\pdfoutput=1 
\documentclass[useAMS,usenatbib,letterpaper]{mn2e}

\usepackage{amsfonts}
\usepackage{amsmath,mathrsfs}
\usepackage{graphicx,float}
\usepackage{epsfig}
\usepackage{color}
\usepackage{float}
\usepackage{epstopdf}
\usepackage{amssymb}
\usepackage{pdflscape}
\usepackage{latexsym}

\usepackage[OT2,T1]{fontenc}
\DeclareSymbolFont{cyrletters}{OT2}{wncyr}{m}{n}
\DeclareMathSymbol{\Sha}{\mathalpha}{cyrletters}{"58}
\DeclareMathSizes{12}{20}{4}{10}

\newcommand{\be}{\begin{equation}}
\newcommand{\ee}{\end{equation}}
\newcommand{\bdm}{\begin{displaymath}}
\newcommand{\edm}{\end{displaymath}}
\newcommand{\bea}{\begin{eqnarray}}
\newcommand{\eea}{\end{eqnarray}}
\newcommand{\nn}{\nonumber}
\newcommand{\kv}{{\bf k}}
\newcommand{\vv}{{\bf v}}

\newcommand{\ff}{{\mathcal F}}
\newcommand{\dif}{{\mathrm d}}

\newcommand{\pa}{\partial}

\newcommand\fnu{\mathsf{f}_\nu^{nr}}

\newcommand{\de}{{\rm d}}

\definecolor{light-gray}{gray}{0.5}

\newcommand{\eq}[1]{eq.\,(\ref{#1})}
\newcommand{\eg}{{\em e.g.}}
\newcommand{\ie}{{\em i.e.}}

\def\camb{\texttt{CAMB}}
\def\class{\texttt{CLASS}}

\def\kMpc{\, h \, {\rm Mpc}^{-1}}

\usepackage[pdftex]{hyperref}

\title[ICs for Accurate Sims with Massive Neutrinos]{Initial Conditions for Accurate N-Body Simulations\\ of Massive Neutrino Cosmologies}
\author[M. Zennaro, {\em et al.}]{
\parbox{\textwidth}{M. Zennaro$^{1,2}$\thanks{E-mail:matteo.zennaro@unimi.it (MZ)}
,
J. Bel$^{2,3}$
,
F. Villaescusa-Navarro$^{4,5,8}$
,\vspace{2mm}\\
C. Carbone$^{2,6}$
,
E. Sefusatti$^{2,4,7}$
,
L. Guzzo$^{1,2}$
}\vspace{3mm}\\
$^{1}$Dipartimento di Fisica, Universit\`a degli Studi di Milano, via Celoria 16, 20133 Milano, Italy\\
$^{2}$INAF - Osservatorio Astronomico di Brera, via E. Bianchi 46, 23807 Merate (LC), Italy\\
$^{3}$Aix Marseille Univ, Univ Toulon, CNRS, CPT, Marseille, France\\ 
$^{4}$INAF - Osservatorio Astronomico di Trieste, Via Tiepolo 11, 34143, Trieste, Italy\\
$^{5}$INFN - Sezione di Trieste, Via Valerio 2, 34127 Trieste, Italy\\
$^{6}$INFN - Sezione di Bologna, Viale Berti Pichat 6/2, 40127 Bologna, Italy\\
$^{7}$INFN - Sezione di Padova, via Marzolo 8, 35131 Padova, Italy\\
$^{8}$Center for Computational Astrophysics, 160 5th Ave, New York, NY, 10010, USA\\}

\begin{document}

\date{\today}


\maketitle

\label{firstpage}

\begin{abstract}
\noindent
The set-up of the initial conditions in cosmological N-body simulations is usually implemented by {\em rescaling} the desired low-redshift linear power spectrum to the required starting redshift consistently with the Newtonian evolution of the simulation. The implementation of this practical solution requires more care in the context of massive neutrino cosmologies, mainly because of the non-trivial scale-dependence of the linear growth that characterises these models. In this work we consider a simple two-fluid, Newtonian approximation for cold dark matter and massive neutrinos perturbations that can reproduce the cold matter linear evolution predicted by Boltzmann codes such as \camb{} or \class{} with a $0.1\%$ accuracy or below for all redshift relevant to nonlinear structure formation. We use this description, in the first place, to quantify the systematic errors induced by several approximations often assumed in numerical simulations, including the typical set-up of the initial conditions for massive neutrino cosmologies adopted in previous works. We then take advantage of the flexibility of this approach to rescale the late-time linear power spectra to the simulation initial redshift, in order to be as consistent as possible with the dynamics of the N-body code and the approximations it assumes. We implement our method in a public code\footnotemark[2] providing the initial displacements and velocities for cold dark matter and neutrino particles that will allow accurate, \ie{} one-percent level, numerical simulations for this cosmological scenario.

\end{abstract}

\begin{keywords}
methods: analytical -- methods: data analysis -- methods: N-body simulations -- methods: numerical -- methods: statistical -- large-scale structure of Universe.
\end{keywords}

\footnotetext[2]{REPS -- rescaled power spectra for initial conditions with massive neutrinos \url{https://github.com/matteozennaro/reps}}

\section{Introduction}

Numerical N-body simulations constitute an essential tool for the interpretation of cosmological observables in redshift as well as weak-lensing surveys. Indeed, a great effort is currently made by several groups to build realistic mocks of the large-scale matter and galaxy distributions \citep[see, \eg][]{AnguloEtal2012, AlimiEtal2012, WatsonEtal2014, KlypinEtal2014, FosalbaEtal2015}. 

In fact, while the evolution of perturbations in the linear regime, {\ie} at large scales and high redshift, is well described by Boltzmann codes such as {\camb} \citep{LewisChallinorLasenby2000} or {\class} \citep{Lesgourgues2011} that accurately reproduce Cosmic Microwave Background (CMB) observations \citep[\eg][]{Planck2015overview}, numerical simulations play a key role in providing predictions for the {\em nonlinear} evolution of structures at small scales and low redshift, where perturbation theory breaks down. 

An important difference between these two fundamental tools in modern cosmology is the fact that Boltzmann codes account for the baryonic and radiation components of the energy density as well as for relativistic effects. N-body cosmological simulations, on the other hand, typically work within the Newtonian approximation, expected to be valid for small density perturbations (weak field) and small velocities, as it is the case for a dominant non-relativistic matter component (Cold Dark Matter or CDM), assuming as an input the expansion history of the Universe as determined by the energy content as a function of redshift. Only fairly recently a certain attention is being devoted to the issue of general relativistic corrections to the initial conditions and to large-scale observables \citep[see \eg][]{ChisariZaldarriaga2011, BruniThomasWands2014, RigopoulosValkenburg2015, ValkenburgHu2015, FidlerEtal2015, HahnParanjape2016} or to the development of a fully general relativistic N-body code \citep[][]{AdamekDurrerKunz2014, AdamekEtal2015,AdamekEtal2016}. Such codes are needed, especially for testing cosmology on scales approaching the horizon, even if Newtonian simulations remain  adequate for the study of sub-horizon scales \citep{JeongEtal2011}.

Furthermore, N-body simulations often take advantage of additional approximations, mainly for practical reasons. For instance,  the contribution to the background expansion of the residual radiation density at the initial redshift is often ignored, although it can amount to a few percent of the total density at $z_i\simeq100$. Possible, related systematic errors are usually avoided, in simulations of $\Lambda$CDM models, by assuming for the power spectrum describing the initial conditions the z=0 output  of a Boltzmann code properly rescaled to the initial redshift by assuming the Newtonian approximation. This practical procedure ensures the recovery of the desired linear theory at low redshift at the expenses of predictivity at high redshift. 

The situation, as one can expect, becomes more complicated in the context of massive neutrino cosmologies. These particles, in fact, are characterised by a Fermi-Dirac momentum distribution with an appreciable tail in the relativistic range, particularly at high redshift. What is more important for clustering studies is that, in these models, the linear growth factors of both cold matter (cold dark matter and baryons) and neutrino perturbations present a characteristic scale-dependence describing the evolution of the power suppression at low scales induced by neutrino free-streaming and such scale dependence also evolves with redshift. For this reason, a rescaling of a low-redshift power spectrum to the initial redshift of the simulation is not trivial and, therefore, simulations of massive neutrino models often assumed the density power spectra for the different species provided by a Boltzmann code {\em at the initial redshift} \citep{VielHaehneltSpringel2010, BirdVielHaehnelt2012, VillaescusaNavarroEtal2013, AliHaimoudBird2013, RossiEtal2014}

However, this procedure can lead to systematic errors, negligible only for sufficiently large neutrino masses and/or sufficiently low values of the initial redshift. In the context of current cosmological investigations, such issues can become relevant. On one hand, small values of neutrino masses are currently favoured \citep[\eg][]{PalanqueDelabrouilleEtal2015}; on the other hand, a low initial redshift might be responsible for other systematic errors related to transients from the initial conditions \citep{Scoccimarro1998, CroccePueblasScoccimarro2006, LHuillierParkKim2014, SchneiderEtal2015}.  

In this work we first quantify the systematics errors resulting from different possible choices for the initial conditions and the background evolution in massive neutrino simulations. If we allow values of the initial redshift as large as $z_i=100$, and require, at the same time, that initial perturbations are determined by a Boltzmann code at that redshift, we need to make sure that the {\em same physical conditions} are reproduced by the simulation in order to avoid unphysical discontinuities. As we will see, this can be achieved with some tweaking of the initial conditions, while failing to do so will result in systematic errors affecting the output power spectra of the simulation. We will consider, in particular, the following issues that could be easily overlooked in the implementation of numerical simulations:
\begin{itemize}
\item the contribution of radiation to the expansion history; 
\item the fraction of massive neutrinos still relativistic and therefore contributing to the radiation energy density (inevitably contributing to the matter energy density in particle-based simulations);
\item the scale-dependence of the {\em growth rate}, $f(k)\equiv\de \ln D(k,a)/\de\ln a$ for the cold matter perturbations in the initial conditions (this could be approximated, for instance, by a constant $f\simeq \Omega_m^{0.55}(z)$, as customary for $\Lambda$CDM cosmologies, moreover equal to unity when radiation is neglected and $\Omega_m=1$);
\item the contribution to the radiation {\em perturbations} of the relativistic fraction of massive neutrino at large redshift, a quantity necessarily equal to zero in particle-based simulations where all neutrinos are always accounted for as non-relativistic  particles that source the gravitational potential;
\item relativistic corrections at near-horizon scales, leading to a scale-dependence of both the growth factor and the growth rate of all particle species (only relevant for very large-volume simulations).
\end{itemize}
One can see that the first two points do not represent a real problem as any radiation contribution can easily be included in the (external) evaluation of the Hubble parameter \citep[as in][]{AliHaimoudBird2013} yet, we will estimate the effect of ignoring radiation in the background anyway. The third point is of interest in $\Lambda$CDM cosmologies only when radiation is indeed accounted for at the initial redshift since otherwise the growth rate $f=1$ in an exact Einsten-de Sitter Universe. For non-vanishing neutrino masses, the growth rate is, moreover, scale-dependent and we will quantify the error resulting from the $f\simeq 1$ approximation. The fourth point remarks that, at the initial redshift, there is still an appreciable tail of relativistic neutrinos that is necessarily ignored in some implementations of massive neutrino simulations. The last point comes from the fact that, in the Newtonian framework often employed in simulations, radiation (photon) perturbations are neglected, leading to inaccuracies in the power spectrum on large scales, as shown by \citet{BrandbygeEtal2016} and \cite{ValkenburgVillaescusa-Navarro2016}.

In order to quantify the effects of the approximations assumed to describe the issues mentioned above, we shall employ a simple two-fluid, Newtonian approximation for cold dark matter and massive neutrinos density perturbations. The solution to the corresponding differential equations can reproduce the linear evolution predicted by Boltzmann codes such as {\camb} or {\class} with a 0.1\% accuracy or below (on the CDM and total matter power spectra) for all redshifts and neutrino masses relevant to nonlinear structure formation \citep{BlasEtal2014}, with the exception, of course, of general-relativistic effects at large scales. 

Furthermore, the two-fluid solution will allow the {\em exact rescaling of the desired low redshift, linear power spectrum to the initial redshift of the simulation in massive neutrino cosmologies in the same fashion similarly to what is usually done in the $\Lambda$CDM case}. We will show that, in this way, we are able to ensure sub-percent errors on the linear evolution of cold dark matter and total matter perturbations in N-body simulation, at any redshift of interest for most cosmological applications. 

So far several different implementations of N-body simulations of massive neutrino cosmologies have been considered in the literature. In the {\em grid-based} method proposed by \citet{BrandbygeHannestad2009} neutrino perturbations are solved in linear theory on a grid. In the {\em particle-based} method employed by \citet{BrandbygeEtal2008}, \citet{VielHaehneltSpringel2010}, \citet{VillaescusaNavarroEtal2013}, \citet{CastorinaEtal2015} and \citet{CarbonePetkovaDolag2016} \citep[but see][for earlier applications of a similar set-up to higher-mass particles]{KlypinEtal1993, PrimackEtal1995} neutrino perturbations are instead described in terms of particles, similarly to CDM, but including a thermal component in their initial velocity. An hybrid version of these two methods has been presented in \citet{BrandbygeHannestad2010}, while an alternative semi-analytic approach is explored by \citet{AliHaimoudBird2013}. \citet{AgarwalFeldman2011} and \citet{UpadhyeEtal2014} consider an implementation purely based on CDM particles, where a fraction of them presents initial perturbations described by neutrino and baryon transfer functions. In this case, however, the correct scale-dependence of the matter power spectrum can only be reproduced at a single redshift. Finally, \citet{BanerjeeDalal2016} presented a method to simulate neutrino cosmologies combining N-body and fluid techniques, aiming at alleviating the shot-noise that affects purely particle simulations.

We will employ the proper scale-dependent rescaling of the linear power spectrum based on the two-fluid approximation, to provide the initial conditions for a set of particle-based simulations. This method has been proved to better capture neutrino effects at the nonlinear level \citep{BirdVielHaehnelt2012, VillaescusaNavarroEtal2013} while allowing studies of the relative velocity among the different species \citep{InmanEtal2015}. However, our results are rather general and can be applied to any method in which the source of the gravitational potential receives contributions from the (linear or nonlinear) neutrino density field, therefore including the \textit{grid-based} approach or the semi-analytic method by \citet{AliHaimoudBird2013}. Such approaches may correctly treat the relativistic neutrino perturbation and can therefore use as an input the linear power spectrum at the initial redshift. Note however that they neglect radiation (photon) perturbations for the evolution of CDM particles, thus requiring rescaling of the $z=0$ power spectrum in order to avoid inaccuracies on large scales (near horizon and above).

This paper is organised as follows. In Section \ref{sec:theory} we introduce massive neutrino models and the main properties of the linear solution to their two-fluid perturbations equations. Section \ref{sec:ICs} shows some possible systematic errors that can arise from different approximations considered in setting the initial conditions of simulations, both with and without massive neutrinos. We quantify the impact of each of them on the linear power spectrum, showing how they can be kept under control. This analysis provides us with a method for rescaling a low redshift power spectrum to the initial redshift. In Section \ref{sec:sims} we test this method on a set of N-body simulations. We present our conclusions in Section \ref{sec:conclusions}.
 
\section{Linear matter perturbations in massive neutrinos cosmologies}
\label{sec:theory}

This section briefly summarises some cosmological effects of a non-vanishing neutrino mass relevant to our purposes. We refer the reader to  \citet{LesgourguesPastor2006} for a comprehensive review. 

Of particular importance for the applications considered in this work is the evolution of the neutrino background density, both in its radiation and matter components, described in the following section. In section \ref{ssec:fluid} instead we will present the two-fluid approximation for the evolution of the cold dark matter and neutrino perturbations.

\subsection{Evolution of massive neutrino density}
\label{ssec:background}

We assume the energy content of the Universe to be composed of cold dark matter (CDM, with total density $\rho_c$ and relative density $\Omega_c$), baryons ($\rho_b$, $\Omega_b$), photons ($\rho_\gamma$, $\Omega_\gamma$), and neutrinos ($\rho_\nu$, $\Omega_\nu$) along with a cosmological constant ($\rho_\Lambda$, $\Omega_\Lambda$). We will refer to the sum of CDM and baryon densities as the total ``cold matter'' component with relative energy density given by 
\be
\Omega_{cb}\equiv\Omega_{c}+\Omega_{b}\,,
\ee
while we will distinguish a massless (relativistic) neutrino component ($\rho_\nu^r$, $\Omega_\nu^r$) from a massive (non-relativistic) component ($\rho_\nu^{nr}$, $\Omega_\nu^{nr}$) such that
\be
\Omega_{\nu}\equiv\Omega_\nu^r+\Omega_\nu^{nr}\,.
\ee
The total matter {\em tout-court} will be given by CDM, baryons and massive neutrinos, and we will denote its relative energy density as
\be
\Omega_{m}\equiv\Omega_{cb}+\Omega_{\nu}^{nr}\,.
\label{omeff}
\ee

The evolution of both photon and neutrino densities depends on their momentum distributions and can be expressed respectively as 
\begin{equation}
\rho_\gamma(z)=\frac{\pi^2}{15}\,(k_B T_{\gamma,0})^4\,(1+z)^4\,,
\label{densityphoton}
\end{equation}
where $k_B=8.617342\times 10^{-5} {\rm eV\,K}^{-1}$ is the Boltzmann's constant ($h$ and $c$ being equal to unity) and, for a neutrino species of mass $m_{\nu,i}$, expressed in eV,
\begin{equation}
\rho_{\nu,i}(z)=\frac{(k_B T_{\nu,0})^4}{\pi^2}\,(1+z)^4\ff\left[ \frac{m_{\nu,i}}{k_B T_{\nu,0}(1+z)}\right],
\label{densityneutrino}
\end{equation}
where $T_{\gamma,0}$ and $T_{\nu,0}$ are respectively the photon and neutrino temperature today, while the function $\ff$ is defined as 
\begin{equation}
\ff(y)\equiv\int_0^\infty\!\!\frac{x^2\sqrt{x^2 + y^2}}{1+e^x}\dif x.
\label{ff}
\end{equation}
It is convenient to express the neutrino energy density, eq.~(\ref{densityneutrino}), in terms of the photon density as 
\begin{equation}
\rho_{\nu,i}(z)=\frac{15}{\pi^4}\,\Gamma_{\nu}^4\,\rho_\gamma(z)\,\ff\left[ \frac{m_{\nu,i}}{k_B T_{\nu,0}(1+z)}\right],
\label{neutrino}
\end{equation}
where $\Gamma_\nu\equiv T_{\nu,0}/T_{\gamma,0}$ is the neutrino to photon temperature ratio today. In the limit of instantaneous decoupling we have 
\begin{equation}
\Gamma_{\nu,\rm inst}=\left( \frac{4}{11}\right)^{1/3}\,,
\end{equation}
while this value must be slightly modified if we want to take into account the fact that the decoupling between photons and neutrinos is not an instantaneous process \citep{HannestadMadsen1995,DolgovHansenSemikoz1997,EspositoEtal2000} and the distortions to the neutrino temperature spectrum introduced by flavour oscillations  \citep{ManganoEtal2005}.  Such corrections are usually expressed in terms of an effective number of neutrino relativistic degrees of freedom defined as 
\begin{equation}
N_{\rm eff} = N_\nu\, \dfrac{\Gamma_\nu^4}{\Gamma_{\nu,\rm inst}^4}\,.
\label{neff}
\end{equation}
We assume the value $N_{\rm eff}=3.046$ corresponding to $\Gamma_\nu=0.71649$ \citep{ManganoEtal2002,ManganoEtal2005}. It should be noted, however, that different approximations in the modelling of the decoupling process lead to variations in the value of $\Gamma_\nu$ with negligible impact on the quantities of interest in this work: for instance, in a cosmology with massless neutrinos, using the instantaneous decoupling value instead of the default one results in a $0.01\%$ difference on the value of the Hubble rate at $z=100$.

In this work we will limit ourselves, for simplicity, to the case of $N_\nu=3$ degenerate massive neutrinos of total mass 
\begin{equation}
M_\nu \equiv \sum_{i=1}^{N_\nu}\,m_{\nu,i}\,.
\end{equation}
Under this assumption, the evolution of the neutrino contribution to the expansion rate of the Universe can be expressed therefore as 
\begin{equation}
\begin{split}
\Omega_\nu(z)\,E^2(z) & = \frac{15}{\pi^4}\,\Gamma_\nu^4\, N_\nu\,\Omega_{\gamma,0} \,(1+z)^4\, \\
& \times \ff\left[ \frac{M_\nu/ (\Gamma_\nu\, N_\nu\, k_B\, T_{\gamma,0}) }{1+z} \right],
\end{split}
\label{onue2}
\end{equation}
where $E(z)$ describes the time dependence of the Hubble rate, such that $H(z)\equiv H_0 E(z)$. 

Eq.~(\ref{onue2}) is the expression we will adopt to describe the neutrino energy density, accounting for both the radiation and matter behaviour at different epochs. The Hubble parameter will therefore be given by
\begin{equation}
\begin{split}
H(z) = H_0 [ & \Omega_{\gamma,0} (1+z)^4 + \Omega_{cb,0}(1+z)^3 + \\
    + & \Omega_\nu (z) E^2(z) + \Omega_\Lambda ]^{1/2}\,,
\end{split}
\label{hofz}
\end{equation}
where $\Omega_{cb,0}$ and $\Omega_{\Lambda,0}$ represent the present cold matter and cosmological constant relative contributions to the energy density. $\Omega_{\gamma,0}$, instead, represents the residual contribution of photons, given by
\begin{equation}
\Omega_{\gamma,0}\,h^2=2.469\times 10^{-5} \,,
\end{equation}
obtained from  \eq{densityphoton} in terms of the CMB temperature, assuming $T_{\gamma,0}=2.7255$ K.\footnote{We remark that in a $\Lambda$CDM cosmology with massless neutrinos, in the computation of the Hubble function expressed as in \eq{hofz}, the neutrino energy density parameter is not given by \eq{onue2} but by its relativistic limit,
\begin{equation}
\Omega_\nu(z)E^2(z) = N_{\rm eff} \frac{7}{8} \left( \frac{4}{11} \right)^{4/3} \Omega_{\gamma,0} (1+z)^4,
\end{equation}
and will therefore contribute, to all effects, to the radiation energy density.}

In the non-relativistic, late-time limit  $m_{\nu,i}\gg T_{\nu,0}(1+z)$, or for $z\ll z_{nr}$ with the redshift of non-relativistic transition $z_{nr}$ estimated as 
\begin{equation}
1+z_{nr} \simeq 1890\frac{m_{\nu,i}}{1 ~{\rm eV}}\,,
\end{equation}
one obtains $\ff\rightarrow y\frac{3}{2}\zeta(3)$, where $\zeta$ is the Riemann zeta function so that
\begin{equation}
\rho_{\nu}(z)=\frac{45}{2\pi^4}\,\zeta(3)\,\frac{\Gamma_{\nu}^4\,\rho_\gamma(z)}{T_{\nu,0}(1+z)}\,M_{\nu}\equiv\,n_{\nu}(z)\,M_{\nu}\,,
\label{neutrino-late}
\end{equation}
$n_{\nu}(z)$ being the neutrino number density. In other words, at late times neutrinos can be assimilated to an additional matter component. 
Dividing eq.~(\ref{neutrino-late}) by the critical density one obtains the well-known expression for the neutrino energy density as a function of the total neutrino mass
\begin{equation}
\Omega_{\nu,0} h^2=\dfrac{M_{\nu}}{93.14 \, \mathrm{eV}}.
\end{equation}

\subsection{Matter perturbations in two-fluid approximation}
\label{ssec:fluid}

The study of perturbations in presence of massive neutrinos dates back to \citet{BondEfstathiouSilk1980} \citep[but see also][]{MaBertschinger1995,Wong2008}. A two-fluid approximation to describe the evolution of coupled cold matter and massive neutrino perturbations has been studied by \citet{ShojiKomatsu2010}. More recently, \citet{BlasEtal2014} considered this approximation to describe the evolution at relative low redshift ($z\ll z_{nr}$) in order to compute perturbative predictions for the subsequent nonlinear evolution. By matching the approximate solution to the exact Boltzmann solution at $z=25$ they recover a $z=0$ linear prediction with an accuracy, at $k=0.1\kMpc$, of 0.1\% and and 1\% respectively for the cold matter and neutrino components. 

We should notice that in practical applications, sub-percent accuracy in the determination of neutrino perturbations is not required. In fact, in the first place, in the expression for the total matter power spectrum
\begin{align}\label{ps_matter}
P_m(k) & =\left(1-\fnu\right)^2 P_{cb}(k)+2\left(1-\fnu\right)\fnu P_{cb,\nu}(k)\nonumber\\
& +\left(\fnu\right)^2 \,P_{\nu}(k)
\end{align}
the contributions of the cross-power spectrum between cold matter and neutrinos, $P_{cb,\nu}(k)$, and of the neutrino power spectrum, $P_{\nu}(k)$,  are suppressed respectively by one and two powers of the massive neutrino fraction 
\be
\fnu(z)\equiv\frac{\Omega_\nu^{nr}(z)}{\Omega_m(z)}\,
\ee
with respect to the contribution of the cold-matter power spectrum $P_{cb}(k)$.  In addition, in particle-based simulations, the initial power spectrum of neutrino particles is usually wiped-out at the first time-step by the effect of thermal velocities and recovered dynamically only at later times.

We now introduce the equations describing the evolution of cold matter and neutrino fluctuations. In our treatment, perturbations in the massive neutrino density will contribute to the gravitational potential and therefore affect the growth of cold matter perturbations. For the cold matter, at linear order, the continuity and Euler equations can be expressed as those of a pressure-less fluid \citep[see, \eg{}][]{BernardeauEtal2002}
\begin{align}
\frac{\pa\delta_{cb}}{\pa t} + \frac{\theta_{cb}}{a} &=  0\,, \label{continuity}\\
\frac{\pa\theta_{cb}}{\pa t} + H\theta_{cb}  &= -\frac1a\nabla^2\phi\,, \label{euler}\,,
\end{align}
where $\delta_{cb}=\delta\rho_{cb}/\bar{\rho}_{cb}$ is the cold matter density contrast and $\theta_{cb}\equiv \nabla \cdot \vv_{cb}$ is the divergence of its peculiar velocity field. Regarding the neutrinos, the two-fluid approximation consists in assuming that neutrino perturbations as well are described just in term of two variables, that is the density and velocity divergence, satisfying the same equations
\begin{align}
\frac{\pa\delta_\nu}{\pa t} + \frac{\theta_\nu}{a} &=  0\,, \label{continuity_nu}\\
\frac{\pa\theta_\nu}{\pa t} + H\theta_\nu  & = \frac{c_{s}^2}{a}\nabla^2\delta_\nu-\frac1a\nabla^2\phi\,, \label{euler_nu}
\end{align}
with the difference that the Euler equation accounts for an effective sound speed $c_s$ given by \citep{BlasEtal2014} 
\begin{equation}
c_s =\frac{\delta p_\nu}{\delta \rho_\nu} \simeq 134.423\, (1+z)\left( \frac{1 \, {\rm eV}}{m_\nu} \right){\rm km}~{\rm s}^{-1}\,.
\end{equation}
The free-streaming scale, or wavenumber $k_{\rm fs}$, characterising neutrino clustering, is directly related to the sound speed as 
\begin{equation}
k_{\rm fs}^2\equiv \frac32\,\Omega_m(a)\frac{a^2\,H^2}{c_s^2}\,.
\label{kfs}
\end{equation}
In the limit where massive neutrinos are non-relativistic, \eq{kfs} reduces to the more familiar expression $k_{\rm fs}^2\simeq 0.83(M_\nu/N_\nu)\Omega_{m,0}/(1+z) ~ \kMpc$.

The set of equations (\ref{continuity})-(\ref{euler_nu}) is then closed by Poisson equation relating the peculiar gravitational potential $\phi$ to the total matter perturbations $\delta\rho_m=\delta\rho_{cb}+\delta\rho_\nu^{nr}$
\be
\nabla^2\phi=\frac32 \,H^2\,\Omega_m a^2\,\delta_m\,.
\label{poisson}
\ee

We look for factorizable solutions of the form, in Fourier space, $\delta_i(a,\kv)=D_i(a)\,\delta_{\kv,i}$ and $\theta_i(a,\kv)=a\,H(a)\,\Theta_i(a)\,\theta_{\kv,i}$, the latter including a useful normalisation. We can rewrite eq.s~(\ref{continuity})-(\ref{euler_nu}) as
\begin{align}
\frac{\partial D_{cb}}{\partial \ln a} & = - \Theta_{cb}\,, \label{coupled-eqs1}\\
\frac{\partial \Theta_{cb}}{\partial \ln a} & = A\,\Theta_{cb} + B\,\left[ \left(1-\fnu\right)D_{cb} + \fnu\,D_\nu \right] \,, \label{coupled-eqs2} \\
\frac{\partial D_\nu}{\partial \ln a}  & = - \Theta_\nu \,,  \label{coupled-eqs3}\\
\frac{\partial \Theta_\nu}{\partial \ln a} & = A\Theta_\nu\! + \! B\!\left[ \left(1\!-\!\fnu\right)\!D_{cb}
+ \left(\fnu \!- \!\frac{k^2}{k_{\rm fs}^{~2}}\right)\!D_\nu \right],\label{coupled-eqs4}
\end{align}
where we introduced the functions
\begin{align}
A(a) & \equiv -\left [ 2+\frac{1}{H^2}\frac{\dif H}{\dif t} \right ]\nn\\
 & = \frac{1}{2}\left[  \Omega_{cb} + 2 \Omega_r -2 \Omega_\Lambda + \Omega_\nu+\Omega_\nu^r -2\right ],
\label{afunc}
\end{align}
and
\begin{equation}
B(a)\equiv -\frac{3}{2} \Omega_m, 
\label{bfunc}
\end{equation}
with each $\Omega_i$ a function of time. Here we have $\Omega_\nu^r$ and $\Omega_m$, the former being the fraction of relativistic neutrinos and the latter the effective matter density parameter defined in \eq{omeff}, in which only the non-relativistic species are accounted for and that determines the source of the gravitational potential in the Newtonian approximation.

To explicitly write these two terms, since we are describing neutrinos as a fluid with an effective pressure $p_\nu$, we can introduce the equation of state  $p_\nu=w_\nu\rho_\nu$ where the parameter $w_\nu(a)$ is a decreasing function of time. Then the relativistic fraction can be written as
\begin{equation}
\Omega_\nu^{r}(a)=   3w_\nu(a) \Omega_\nu(a),
\label{onur}
\end{equation}
and the effective matter density parameter can be written as
\begin{equation}
\Omega_m(a) = \Omega_{cb}(a) + [1-3w_\nu(a)]\Omega_\nu(a).
\label{omeffexplicit}
\end{equation}
We can obtain the time-dependence of $w_\nu(a)$ from the energy density scaling $\rho_\nu(a)\sim a^{-3[1+w_\nu(a)]}$ as
\begin{equation}
\begin{split}
3(1+w_\nu)&= - \frac{\dif\ln\rho_\nu}{\dif\ln a} \\
&=4 - y  \frac{\dif\ln\ff(y)}{\dif y} \,,
\end{split}
\label{pression}
\end{equation}
where $y=M_\nu/ [\Gamma_\nu\, N_\nu\, k_B\, T_{\gamma,0}(1+z)]$.
We notice that the derivative of $\ff$ can be expressed in terms of the following integral
$$
\frac{\dif \ff(y)}{\dif y}=y \int_{0}^\infty\frac{u^2}{1+e^u}\frac{\dif u}{\sqrt{u^2+y^2}}.
$$

Finally, we need to provide the boundary conditions. These are given in terms of the initial values for the growth rate of the cold matter component, 
\be
f_{cb} (a)\equiv \frac{\dif \ln D_{cb}(a)}{\dif \ln a}\,
\label{f_cb}
\ee
and of neutrinos
\be
f_{\nu} (a)\equiv \frac{\dif \ln D_{\nu}(a)}{\dif \ln a}\,
\label{f_nu}
\ee
along with the ratio between neutrino and cold matter perturbations 
\be
\beta(a)\equiv \frac{D_\nu(a)}{D_{cb}(a)}\,.
\ee
We find four independent solutions that we linearly combine by imposing the conditions
\begin{equation}
\begin{split}
\mathrm{at}~z = 0_{\phantom{i}} \qquad &\left(1-\fnu\right)D_{cb} + \fnu\,D_\nu \equiv  D_m \equiv 1 , \\ 
\mathrm{at}~z = z_i \qquad &\begin{cases}
\beta D_{cb} - D_\nu = & 0 ,  \\
\Theta_{cb} + f_{cb} D_{cb}  =  & 0,  \\
\Theta_\nu + f_\nu D_\nu  = & 0 ,
\end{cases}
\end{split}
\label{boundary}
\end{equation}
where we have set the amplitude of the total matter growth factor at $z=0$ equal to unity, since we will be interested in rescaling the power spectrum at that redshift.

\subsection{Comparison with Boltzmann codes}
\label{ssec:Boltzmann}

\begin{figure*}
\centering
\includegraphics[width=.98\textwidth]{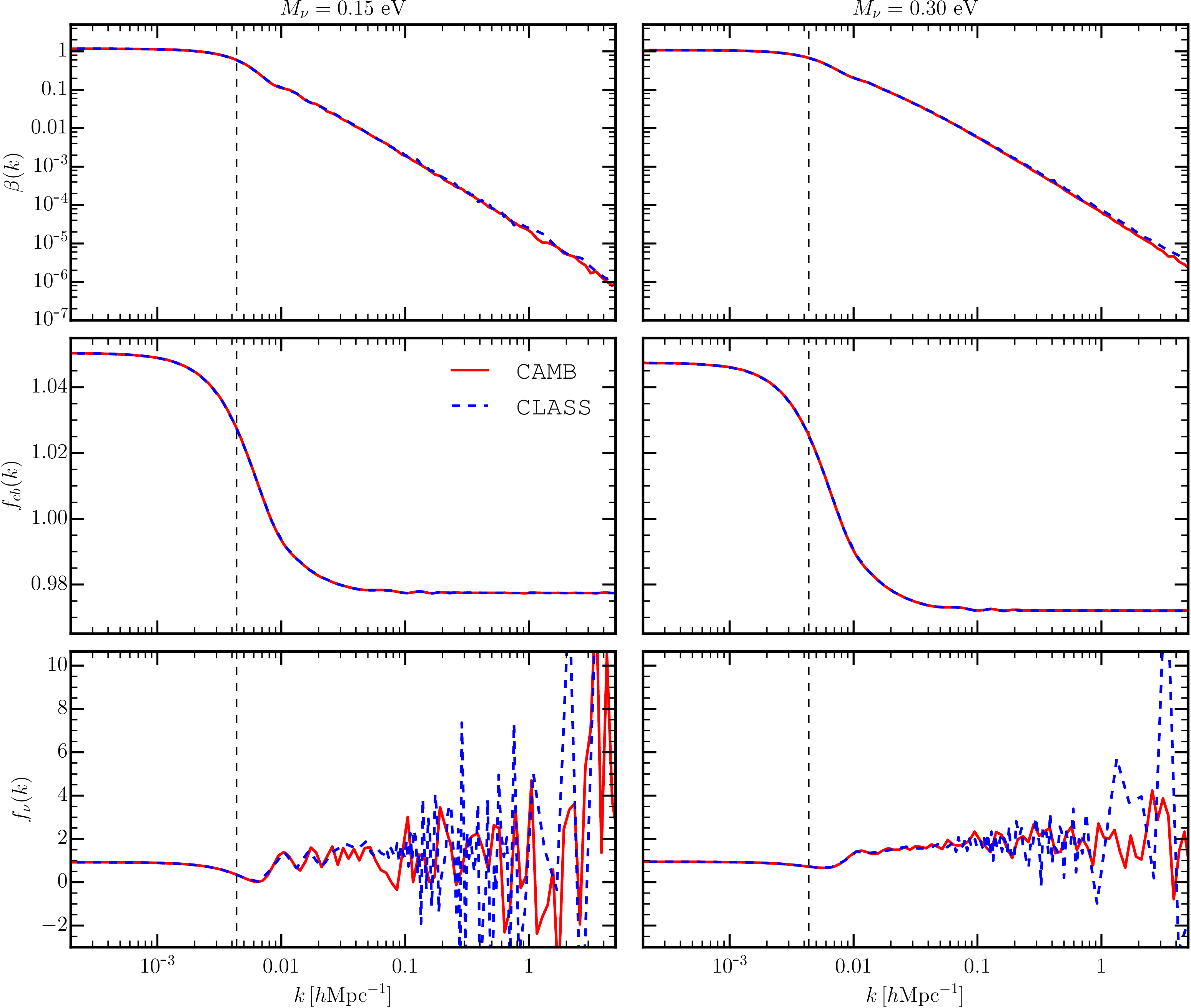}
\caption{Comparison of the initial conditions provided by \camb{} (solid, red curves) and \class{} (dashed, blue curves) at the reference initial redshift $z_i=99$ as a function of scale. {\it Top: } Ratio between CDM+baryon and neutrino fluctuations, $\beta\equiv D_\nu/D_{cb}$. {\it Middle:} CDM+baryon growth rate, $f_{cb}\equiv \dif\ln D_{cb}/ \dif\ln a$. {\it Bottom: } neutrino growth rate, $f_\nu\equiv \dif\ln D_\nu/\dif\ln a$. In the two columns we represented the initial conditions for neutrino masses $M_\nu = 0.15$ (left) and $0.30$ eV (right). The dashed vertical line marks the scale of the horizon at the chosen redshift. }
\label{ic}
\end{figure*}

In this section we compare the two-fluid approximation with the Boltzmann codes \camb{} and \class{}.  

To this end, we shall use as boundary conditions the outputs of these codes at the chosen initial redshift. We therefore start by comparing the initial conditions at $z_i=99$ provided by \camb{} and \class{} in Fig. (\ref{ic}) for two different values of the total neutrino mass and find no significant difference between the two cosmologies. 

\begin{figure*}
\centering
\includegraphics[width=.98\textwidth]{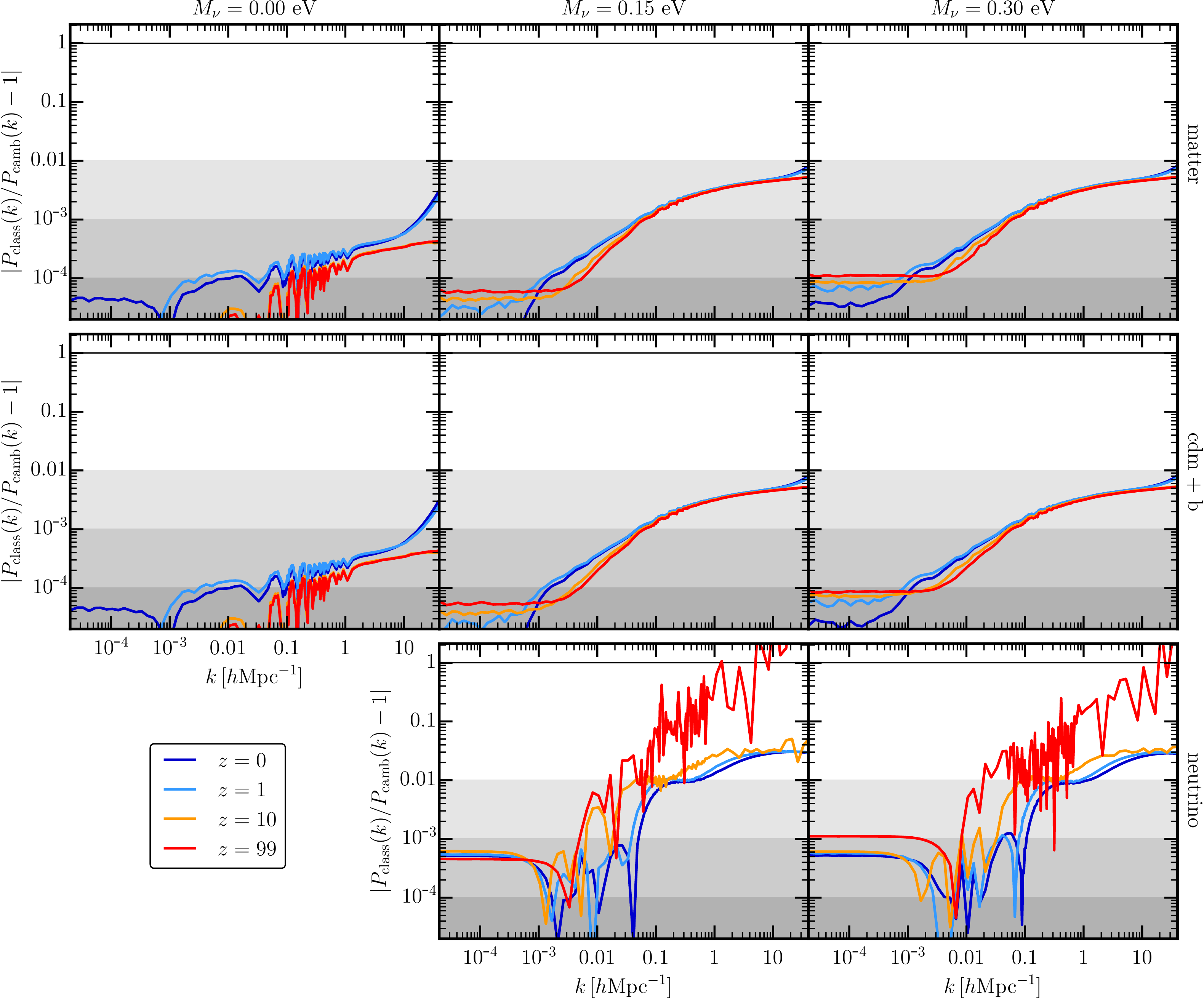}
\caption{Comparison between power spectra from {\camb} and {\class}, $|P_{\rm class}(k,z)/P_{\rm camb}(k,z) - 1|$. The precision parameters of the two codes have been set as in \href{http://arxiv.org/abs/1104.2934}{the comparison paper} between \class{} and \camb{} (for $\Lambda$CDM). The two codes agree on the CDM and total matter power spectra at $10^{-4}$ level at almost all scales in the $\Lambda$CDM case, and their agreement is better than $\sim 0.2\%$ for $k<1h^{-1}$ Mpc when massive neutrinos are included.}
\label{fig:CLASSvsCAMB} 
\end{figure*}

To make sure that our input parameters for the two Boltzmann codes represent exactly the same cosmology, we also check to what extent the power spectra generated with {\camb} and {\class} agree with each other. This is shown in Fig.~\ref{fig:CLASSvsCAMB} as the absolute value of the relative difference evaluated for a $\Lambda$CDM cosmology and two massive neutrino cosmologies at redshifts $z=0,1,10,99$. We adopted the settings suggested in \citet{Lesgourgues2011} and in \citet{LesgourguesTram2011}, where the reference comparison between the two codes has been originally presented. The agreement in the $\Lambda$CDM case is better that $10^{-4}$, while in the massive neutrino cases is better than $0.2$\% for $k<1h$ Mpc$^{-1}$, a result consistent with what indicated by the two reference papers. For the neutrino power spectrum, we find an agreement between the two codes better than $1-2\%$ at all scales below $k \sim 0.1h~\mathrm{Mpc}^{-1}$.

\begin{figure*}
\centering
\includegraphics[width=.98\textwidth]{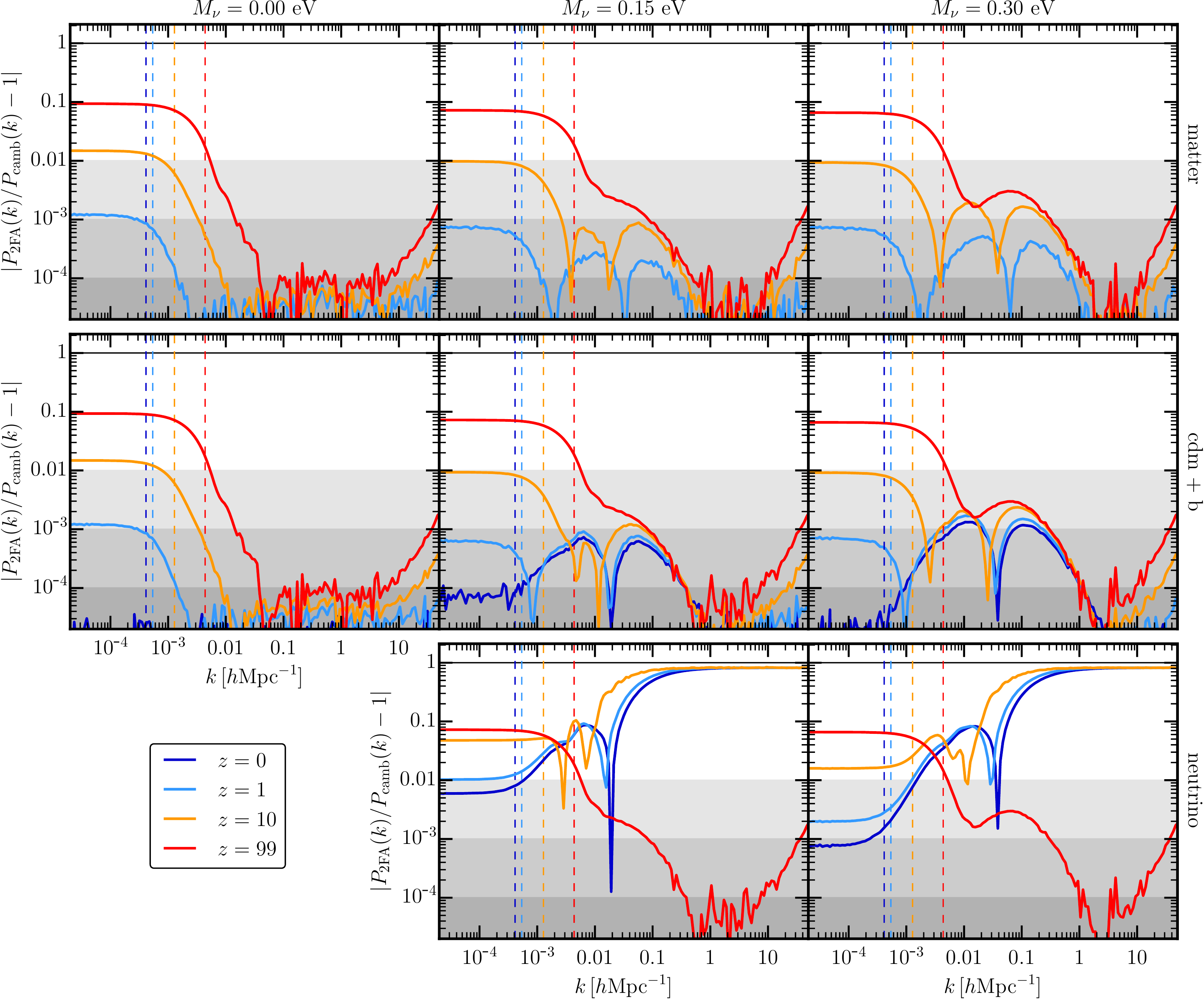}
\caption{Relative difference, $|P_{\rm 2FA}(k,z)/P_{\mathrm{camb}}(k,z)-1|$, between the linear power spectrum $P_{\rm 2FA}(k,z)$ obtained in the two-fluid approximation and the solution from {\camb}, $P_{\mathrm{camb}}(k,z)$. The {\em total} matter $P_{\rm 2FA}$ is normalised to match the \camb{} results at $z=0$.  Upper, middle and lower panels show, respectively, the results for the total matter fluctuations ($P_m$), CDM+b fluctuations ($P_{cb}$) and the neutrinos fluctuations ($P_\nu$). We show three different cosmologies, on the left a standard $\Lambda$CDM one, in the middle column a cosmology with $M_\nu=0.15$ eV and on the right $M_\nu=0.30$ eV. Colours from red to blue denote, in the order, redshifts $z=99,\, 10,\, 1$ and $0$. The dashed horizontal lines mark the scale of the horizon at the corresponding redshift. }
\label{fig:CAMBvs2F} 
\end{figure*}

In Fig.~\ref{fig:CAMBvs2F} we show the comparison between the same linear power spectra computed at the same redshifts using our two-fluid approximation against \camb{}, matching to the Boltzmann code at $z_i=99$. In this case, we normalise our results to agree with the \camb{} {\em total} matter power spectrum at $z=0$, and study any discrepancy might arise at larger redshift or for the other two components even at $z=0$. We have repeated the same analysis using the code \class{}, and all the following results hold for both the codes. Overall, the two-fluid approximation is able to recover the output from the two Boltzmann codes with a precision of a few $10^{-4}$ on the relevant range of scales. In particular in both cases for the total matter and the CDM power spectra we can recognise three different regions:
\begin{enumerate}
\item On very large scales, above the horizon at each redshift, we can see the mismatch induced by our Newtonian approximation, in which photon perturbations are completely neglected, with respect to the relativistic solution provided by the Boltzmann codes. This difference reaches an amplitude of $\sim 10\%$ at $z=99$, maximal given our choice to normalise at $z=0$. 
\item On intermediate scales, within the horizon but for modes below $\sim 10 h \, {\rm Mpc}^{-1}$, our two-fluid approximation agrees very well with both codes (the percentage difference being around 0.01-0.02\%).
\item On very small scales, $k>10h \, {\rm Mpc}^{-1}$, the two fluid approximation breaks down and the discrepancy with the two Boltzmann codes begins growing increasingly. This is expected as the approximate, effective pressure term in the neutrino Euler equation becomes more relevant at large $k$. Nonetheless, the two-fluid approximation is still quite reliable up to $k \sim 50 h \, {\rm Mpc}^{-1}$, the relative difference with the Boltzmann codes being still below $0.1\%$.
\end{enumerate}

We note that, the agreement found for the cold and total matter components between our approximation and the Boltzmann codes, is not present for the neutrino component. In particular, at $k \sim 0.1h~\mathrm{Mpc}^{-1}$ the neutrino power spectrum differs from both \camb{} and \class{} by $\sim 100 \%$. In fact the agreement with the neutrino power spectrum largely depends on the version and choice of precision parameters of the Boltzmann codes \footnote{With the 2014 version of \class{} and the precision parameters therein contained, we are able to obtain an agreement of $1-2\%$ between the two-fluid approximation and the Boltzmann solution \citep[as in][]{BlasEtal2014}, which we do not recover with the latest version of this code.}. Yet, this is not of great importance as no observable or relevant quantity is exclusively dependent on the neutrino perturbations, that only provide, for instance, sub-percent corrections to the total matter power spectrum. Moreover, the solution at the initial redshift (which is the one used for setting the initial conditions for simulations) is always correct and shares the same level of accuracy as the CDM component, since discrepancies arise only in the the evolution of the initial power spectrum and grow with time.

\section{Potential Systematic Errors from the Initial Conditions}
\label{sec:ICs}

The usefulness of the two-fluid approximation consists in allowing us to quickly quantify the possible systematic errors resulting from different assumptions on the initial conditions and the background expansion in simulations. The differential equations presented in the previous section have the advantage, with respect to Boltzmann codes, of reproducing the Newtonian physics simulated by N-Body codes and can therefore mimic the expected evolution of perturbations at the linear level.   

We consider several potentially incorrect approximations, including some affecting as well $\Lambda$CDM cosmologies (\ie{} with massless neutrinos). For instance, one of the most common approximations is to ignore the contribution of the radiation density to the Hubble expansion. This is not a problem in a $\Lambda$CDM cosmology as long as the initial power spectrum is rescaled from a the low-redshift power spectrum consistently with the expansion history assumed for the simulation. However, if we set-up the initial conditions with a linear power spectrum computed by a Boltzmann code {\em at the initial redshift} we should then assume the correct Hubble expansion including the radiation contribution. Doing otherwise, as we will see, can result in a systematic error on the linear growth at low redshift of a few percent. Other approximations might have lesser consequences but the related errors can cumulate and should nevertheless be taken into consideration. 

The reference growth factors we use for performing our comparison are computed solving our system of fluid equations in the optimal set-up, {\em within the limitations imposed by the Newtonian approximation}. This means we use the neutrino density and effective pressure computed from the momentum distribution function as in equations \ref{onue2} and \ref{pression}, but we are not accounting here for any general-relativistic correction at scales approaching the horizon.

By solving the coupled equations~(\ref{coupled-eqs1})-(\ref{coupled-eqs4}) we obtain the growth factor of each species, normalised to have the same value at the initial redshift $z_i=99$ in order to quantify the effect at $z=0$ in terms of the expected error on the power spectrum, therefore as the ratios
\begin{displaymath}
\frac{D^2_{cb}(k,z=0)}{D^2_{cb,\,{\rm ref}}(k,z=0)}\,,\qquad \frac{D^2_{\nu}(k,z=0)}{D^2_{\nu,\,{\rm ref}}(k,z=0)}\,,
\end{displaymath}
while the total matter growth factor  $D_{m}(k,z)$ is obtained as   
\be
D_{m}(k,z)=(1-\fnu) \,D_{cb}(k,z)+\fnu\,D_{\nu}(k,z)\,. 
\ee
for both the reference and approximated solutions.

\begin{figure*}
\centering
\includegraphics[width=0.98\textwidth]{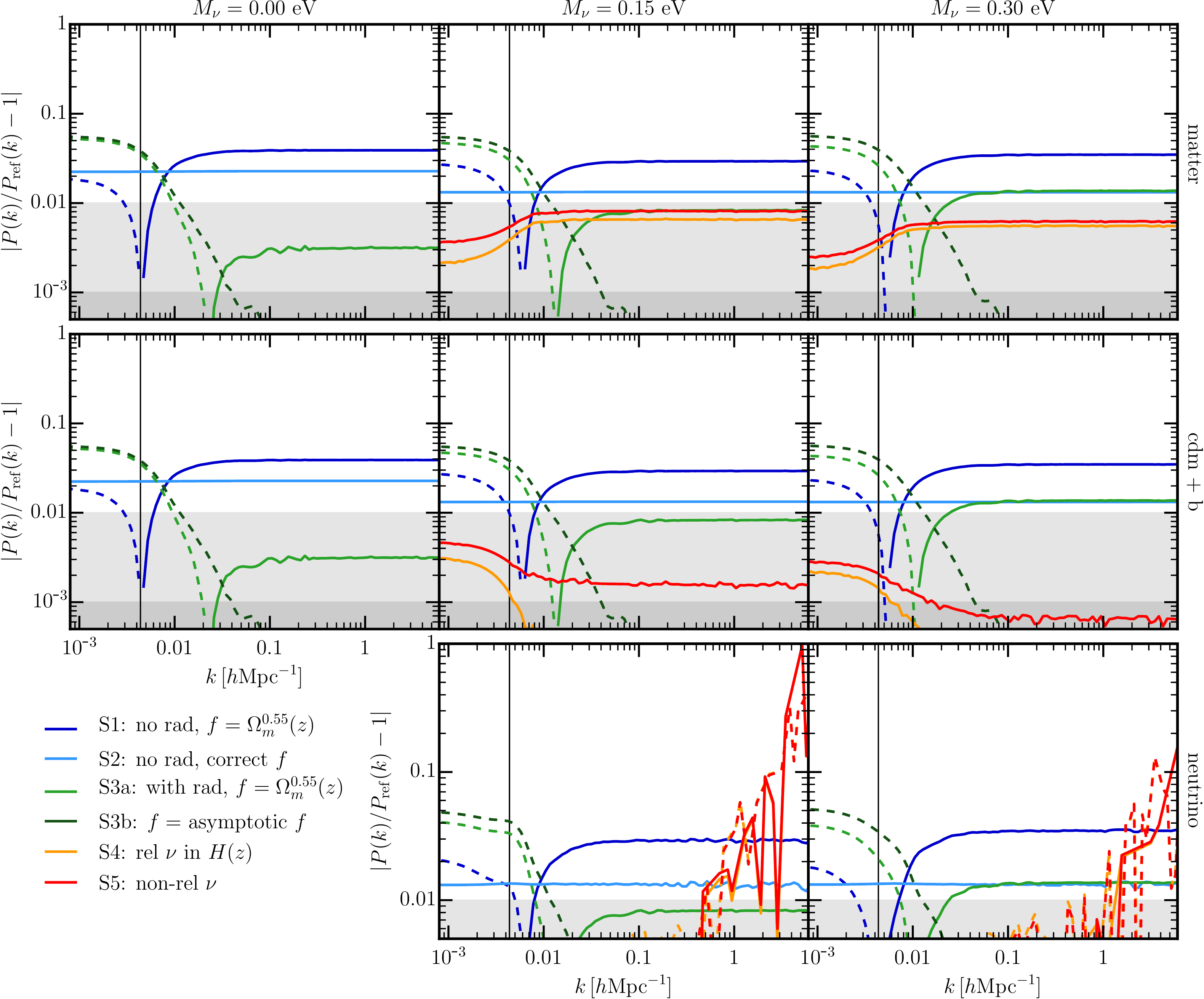}
\caption{The impact of five common approximations assumed in simulations on the linear power spectrum at $z=0$ with respect to a reference power spectrum. The vertical gray line marks the scale of the horizon at $z_i = 99$. The reference power spectrum is the one computed with the two-fluid approximation, with no further approximation (\ie{} it is the power spectrum shown in figure \ref{fig:CAMBvs2F}, that agrees with the output of the Boltzmann codes apart from the fact of being in Newtonian limit). \textit{Blue lines} show the effect of neglecting photons in the background in combination with using the $\Lambda$CDM approximated growth rate $f=\Omega_m^{0.55} \simeq 1$ (scenario S1). \textit{Light blue lines} show the impact on the low redshift power spectrum of neglecting photons but using the correct, scale dependent growth rate (scenario S2). Note that, in the left panel, besides photons, we do not take into account massless (hence relativistic) neutrinos. If we do the contrary and we include photons in the background but use the $\Lambda$CDM parametrization for the growth rate (scenario S3a) we obtain the \textit{light green lines}, while a scenario with scale-independent growth rates where the value is the correct one within the horizon is shown as scenario S3b (dark green lines). \textit{Orange lines} show the impact of including the relativistic fraction of neutrinos in the background, but not in the computation of the peculiar gravitational potential, an unavoidable approximation in particle-based simulations where neutrinos are implemented as particles with constant mass (S4). Finally, \textit{red lines} refer to S5, where we treat neutrinos as a completely non-relativistic specie.}
\label{fig:SUs}
\end{figure*}

\begin{table*}
\centering
\begin{tabular}{l c c c c }
Scenario & $\Omega_\gamma$ & $\Omega_\nu^r$ &  $f=\de\ln D/\de\ln a$ & $B(a)$ \\
\hline
S1: No photons, $f$ constant            & $0$ & $0$ & $\Omega_m^{0.55}\simeq 1$ & correct \\
S2: No photons + $f(k)$ & $0$ & $0$ & correct & correct \\
S3a: $f$ constant, $f=\Omega_m^{0.55}$ & correct & correct & $\Omega_m^{0.55}$ & correct \\
S3b: $f$ constant, asymptotic value & correct & correct & asymptotic & correct \\
S4: Constant mass neutrino particles & correct & correct & correct & $\Omega_{m} = (\Omega_{cb,0}+\Omega_{\nu,0})a^{-3}$ \\
S5: No relativistic neutrinos & correct & $0$ & correct & $\Omega_{m} = (\Omega_{cb,0}+\Omega_{\nu,0})a^{-3}$ \\
\hline
\end{tabular}
\caption{Different scenarios of the considered, possible approximations in the initial conditions and in the Hubble expansion. $B(a)$ is the source term of the gravitational potential in the Newtonian approximation defined in \eq{bfunc}. When it is correct, $\Omega_{m}$ includes only the actual non-relativistic fraction of neutrinos, following \eq{omeffexplicit}; otherwise, \textit{all} neutrino perturbations are considered as non-relativistic (irrespective of possible relativistic tails) and always act as sources of gravity.}
\label{tab:sims}
\end{table*}

We will quantify the systematic error on the linear growth of total and cold matter and neutrino perturbations as a function of redshift and scale for five different scenarios. These are summarised in Table~\ref{tab:sims} while we show our numerical results in Fig.~\ref{fig:SUs}. In what follows we describe each scenario, discussing the resulting systematic error.

\subsubsection*{S1: no radiation and constant, initial growth rates}  

In the first scenario we neglect the radiation contribution to the Hubble expansion setting
\begin{equation}
\Omega_{\gamma,0} = 0,
\end{equation}
in \eq{hofz} and we impose the constant values for the growth rate
\begin{equation}
f_{cb}=f_\nu=\Omega_m^{0.55}(z_i)\simeq 1
\end{equation}
in the initial conditions, \eq{boundary}, at $z_i=99$ (we remark that even a large error on the neutrino growth rate, $f_\nu$ has negligible consequences). The combination of these two assumptions is due to the second being a consequence of the first, for $\Lambda$CDM cosmologies, since $\Omega_m(z_i)\simeq 1$ when no radiation is present.  In massive neutrino cosmologies this is not strictly true because of the scale-dependence of the growth rate and it therefore represents a further approximation for these models.

In general a larger density of relativistic species leads to a smaller growth of matter fluctuations. For this reason we expect a higher amplitude of the power spectrum when we neglect the contribution of radiation to the background. This indeed is what we see in Fig.~\ref{fig:SUs} for S1, where the amplitude of the power spectrum at all scales that were within the horizon at $z_i$, is larger than the reference case at $z=0$ by $3-4 \%$. Moreover, since we are neglecting the scale-dependence of the growth rate in the initial conditions, we see that the error induced presents, in turn, a peculiar scale-dependence  at large but still observable scales.

\subsubsection*{S2: no radiation}  

In the second scenario we consider the case of using the correct growth rate even when there is no radiation in the background; to this purpose, we compute the Hubble function setting
\begin{equation}
\Omega_{\gamma,0} = 0,
\end{equation}
in \eq{hofz}. The growth rates are computed as numerical derivatives of the power spectrum at the initial redshift (\ie{} numerically solving eq. (\ref{f_cb}-\ref{f_nu}) on each scale). In this case the resulting error is clearly scale-independent but still corresponds to more than 2\% on the low redshift $\Lambda$CDM matter power spectrum, reduced to around 1.5\% in the massive neutrino case, due to the different effective number of relativistic neutrinos contributing to radiation.

\subsubsection*{S3:  constant growth rate, $f$ } 

In the third scenario we overturn the situation and consider the effect of using the correct background evolution (with radiation and with the proper contribution of relativistic neutrinos, as in \eq{hofz}) keeping the growth rate fixed (\ie{} scale independent). In one case, that we call S3a, the constant value of the growth rate is given by the approximation
\be
f_{cb}=f_\nu=\Omega_m^{0.55}(z_i)\,,\qquad {\rm scale-independent}.
\ee
which is valid in a $\Lambda$CDM cosmology with no radiation in the background (though we do have radiation in the background). Here the approximation has actually three implications: \textit{(i)} the scale independence does not account for relativistic effects on large scales, such as the contribution of radiation perturbations, \textit{(ii)} the scale independence does not account for the suppression induced by neutrinos at small scales, \textit{(iii)} this approximate value is valid only with no radiation in the background. The main discrepancy from the correct low redshift power spectrum appears at near-horizon scales, which are affected by about $5-6\%$. On the contrary, for scales within the horizon this approximation results in a discrepancy of $\sim0.3\%$ in the $\Lambda$CDM case, but becomes more pronounced, {\em and relevant}, being above the percent level, in presence of massive neutrinos. 

It is possible, of course, to consider a similar scenario in which the growth rates are again scale-independent, but they assume the correct asymptotic value of the cold matter growth rate within the horizon. To do so we numerically compute the growth rate as in \eq{f_cb}, considering only the asymptotic value towards small scales (therefore, well within the horizon). We note that for the $\Lambda$CDM case, this procedure corresponds to the approximation $f \simeq \Omega_m^{0.667}$. In this case, that we name S3b, in the evolved $z=0$ power spectrum, only scales above horizon show a discrepancy with respect to the Boltzmann solution (of the same amplitude as in S3a), while scales within the horizon agree at 0.01\% level. The slight difference between case S3a and S3b for scales above horizon in the massive neutrino cosmologies is due to the residual systematic error induced by the fact that the asymptotic value for neutrinos ignores their scale dependence.

\subsubsection*{S4: constant mass neutrino particles} 

This approximation, as the following one, is specific to massive neutrinos models. In fact, at redshifts as high as $z_i\simeq 100$, there is still a significant tail of relativistic neutrinos that does not contribute to the gravitational potential. Particle-based N-body simulations, however, assume massive neutrinos to be non-relativistic (\ie{} matter) at all redshifts $z\le z_i$. Avoiding this approximation would require to allow the mass of neutrino particles to vary in time. To the best of our knowledge, however, no code in the literature considered this possibility.

To reproduce this scenario, therefore, we modify the function $B(a)$ defined in \eq{bfunc}, which is the source of gravity in the Newtonian approximation,
\begin{equation}
B(a) = -\frac{3}{2} \Omega_m(a), 
\end{equation}
using 
\begin{equation}
\Omega_m (a) = (\Omega_{cb,0}+\Omega_{\nu,0})~a^{-3}
\end{equation}
instead of \eq{omeffexplicit}. This means that in computing neutrino overdensities all neutrinos are treated as non-relativistic particles.
On the other hand, we do not modify the Hubble rate $H(a)$ and the function $A(a)$ (which only depends on the Hubble rate and its first derivative), defined in \eq{hofz} and \eq{afunc} respectively, in order to fully account for relativistic neutrinos in the background.

This approximation results in a negligible error on the power spectrum of the CDM component at low redshift. The effect is larger on the neutrino component, and therefore on the total matter power spectrum. Nonetheless, as neutrinos weight considerably less that CDM, even on the total matter power spectrum the effect is sub-percent.

\subsubsection*{S5: no relativistic neutrinos} 

Finally we consider {\em all } effects of neglecting the relativistic neutrino fraction, that is both on the perturbations (Poisson equation) as on the Hubble expansion, treating neutrinos as a non-relativistic species \textit{also} in the background, as usually done in the literature. To do so, we extend the approximations described in S4 to the computation of the Hubble rate and of the function $A(a)$ (eq. \ref{afunc}). This means that in this case we always use
\begin{equation}
\Omega_\nu (a) = \Omega_{\nu,0} ~ a^{-3}
\end{equation}
and
\begin{equation}
w_\nu(a)=0.
\end{equation}
The resulting error is only slightly larger than the one of S4, however, our  solution is affected by numerical instabilities in the neutrino sector of the coupled differential equations.

\bigskip

From this exercise we conclude that, on the scales that at $z=z_i$ were within the horizon, the greater impact ($>1\%$) comes from neglecting the scale dependence in the initial growth rate of the cold matter perturbations and neglecting the radiation contribution to the Hubble function. We should notice, in addition, that  their combined effect can sum up to an appreciable level, even when individual errors are sub-percent.

\section{Initial conditions for accurate N-body simulation} 

\subsection{Linear rescaling}

A rather obvious but important application of the two-fluid approximation is the proper, scale dependent rescaling of a desired low-redshift power spectrum to the initial redshift of a simulation. As shown in the previous section, this is possible in a fashion completely consistent with the dynamics and approximations assumed by the simulation itself.

The initial power spectrum for the cold matter perturbations would then be obtained as
\be
P_{cb}(k,z_i)=\frac{D^2_{cb,2FA}(k,z_i)}{D^2_{cb,2FA}(k,0)}P_{cb}^{B}(k,0)
\ee 
where $D_{cb,2FA}(z)$ is the growth factor obtained in the two-fluid approximation while  $P_{cb}^{B}(k,0)$ is the desired linear power spectrum from a Boltzmann code at $z=0$. In a similar way one can obtain the initial power spectrum for the neutrinos, although its accuracy is far less important.  

\begin{figure}
\centering
\includegraphics[width=0.48\textwidth]{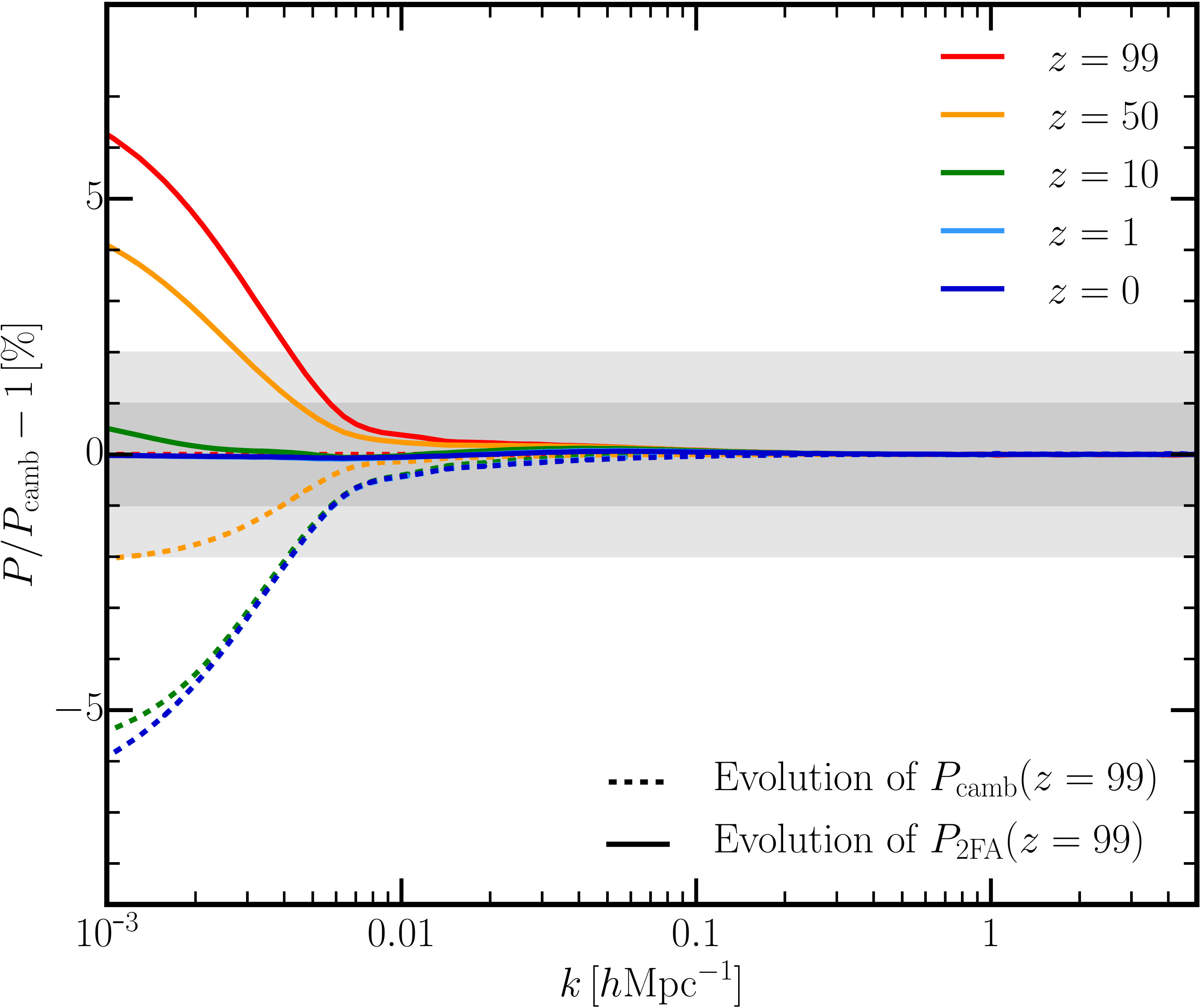}
\caption{We compare the evolution of two different initial linear power spectra at $z=99$: one is directly taken from \camb{}, the other one is obtained with our method, \ie{} the $z=0$ power spectrum from \camb{} has been rescaled to the initial redshift with our two-fluid approximation. The two spectra are then evolved forward using the two-fluid approximation, thus mimicking the linear (Newtonian) evolution in a simulation. In the first case, dashed lines, by construction at high redshift the evolved power spectrum coincides with that from the Boltzmann code, but differs from it at low redshift (reaching a lack of power $>5\%$ at large scales at $z=0$). On the other hand, our rescaling (solid lines) introduces a discrepancy at high redshift and large scales (due to the the fact that in our Newtonian approximation all photon perturbations are inevitably neglected), but allows us to recover a sub-percent agreement with the linear power spectrum from a Boltzmann code at lower redshifts ($z<10$).}
\label{fig:norm}
\end{figure}

In Fig.~\ref{fig:norm} we compare the set-up of the initial conditions we are proposing to the method often applied in previous works. To do so, we fix the initial power spectrum, at $z=99$, in two different ways: the first one is obtained by \textit{directly} setting the \camb{} output at $z=99$ as the initial power spectrum; the second one, which corresponds to our approach, is obtained by rescaling the $z=0$ power spectrum from \camb{} (but the same holds for \class{}) to the initial redshift $z_i$ with the two-fluid approximation. We then evolve these two power spectra using our two-fluid approximation scheme to mimic the linear evolution in a simulation, and compare the different outcomes at lower redshifts. Notice that the  initial growth rate is the correct one in both cases.

In the first case (dashed lines in the plot) the power spectrum, by construction, coincides with the Boltzmann solution at $z_i$, but then shows a lack of power on large scales as we move towards lower redshifts. This is due to the fact that most simulations -- and our two-fluid approximation -- work in a Newtonian framework that cannot account for radiation (photon) perturbations, and therefore cannot correctly reproduce the scale dependent growth of perturbations on large scales. Such discrepancy can be as large as $5-6\%$ and especially affects scales with $k \lesssim 0.01 ~h \mathrm{Mpc}^{-1}$, where we would expect to recover linear predictions with great precision. On the other hand, in the second case (solid lines in figure \ref{fig:norm}), we inevitably loose accuracy at very high redshift, but we are able to recover a sub-percent agreement for $z<10$.

The rescaling therefore introduces a spurious (but motivated) excess of power at large scales and high redshift, that allows us to recover the correct power spectrum at lower redshifts. We note that, since nonlinearities  introduce a coupling between large and small scales, an excess of power on the large scales could in principle alter the nonlinear evolution; however, this potential error is very small since the excess of power is confined to superhorizon modes and high redshifts and, in the next session, we will show that this method allows us to recover with sub-percent accuracy the expected nonlinear power at redshifts below 10.

\subsection{Test with N-body simulations}
\label{sec:sims}

In order to validate our method we have run  N-body simulations with massless and massive neutrinos where the initial conditions have been generated using the method described above. We rescale a $z=0$ power spectrum generated with \camb{} to the initial redshift of the simulation in the two scenarios presented above that can mimic the dynamics in the simulations, namely S4 (in which the relativistic fraction of neutrinos is taken into account in the Hubble function, but not in the computation of the peculiar gravitational potential) and S5 (in which also in the background evolution we treat all neutrinos as a non-relativistic species).

The simulations have been run using the TreePM-SPH code {\sc GADGET-III} \citep{Springel2005}. The size of the periodic simulation box in all our simulations is set to $2$ $h^{-1}{\rm Gpc}$. We have run simulations for three different cosmological models: a massless neutrino cosmology and two models with massive neutrinos corresponding to $M_\nu=0.15$ eV and $M_\nu=0.3$ eV.  The relatively large values of $M_\nu$ are justified by the purpose to test our method. All simulations share the value of the following cosmological parameters: $\Omega_{m}=\Omega_{cb}+\Omega_\nu=0.3175$, $\Omega_{b}=0.049$, $\Omega_{\Lambda}=0.6825$, $h=0.6711$, $n_s=0.9624$ and $A_s=2.13\times10^{-9}$. In the models with massive neutrinos we set $\Omega_{c}=\Omega_{m}-\Omega_\nu$, where $\Omega_\nu=M_\nu/(93.14~{\rm eV}\,h^2)$. We notice that, since all models have the same normalization of the amplitude of the linear power spectrum at the epoch of the CMB, the value of $\sigma_8$ will be different in each model: $\sigma_8=0.834,~0.801,~0.760$ for the models with $M_\nu=0.0,~0.15,~0.30$ eV, respectively.

We follow the evolution of $768^3$ CDM particles, plus $768^3$ neutrino particles in the models with massive neutrinos, from $z=99$ down to $z=0$. In order to carry out convergence tests, we have also run a second set of simulations with $512^3$ CDM and neutrino particles. The gravitational softening length is set to $1/40$ of the mean inter-particle distance, both for CDM and neutrino particles.

\begin{figure*}
\centering
\includegraphics[width=0.98\textwidth]{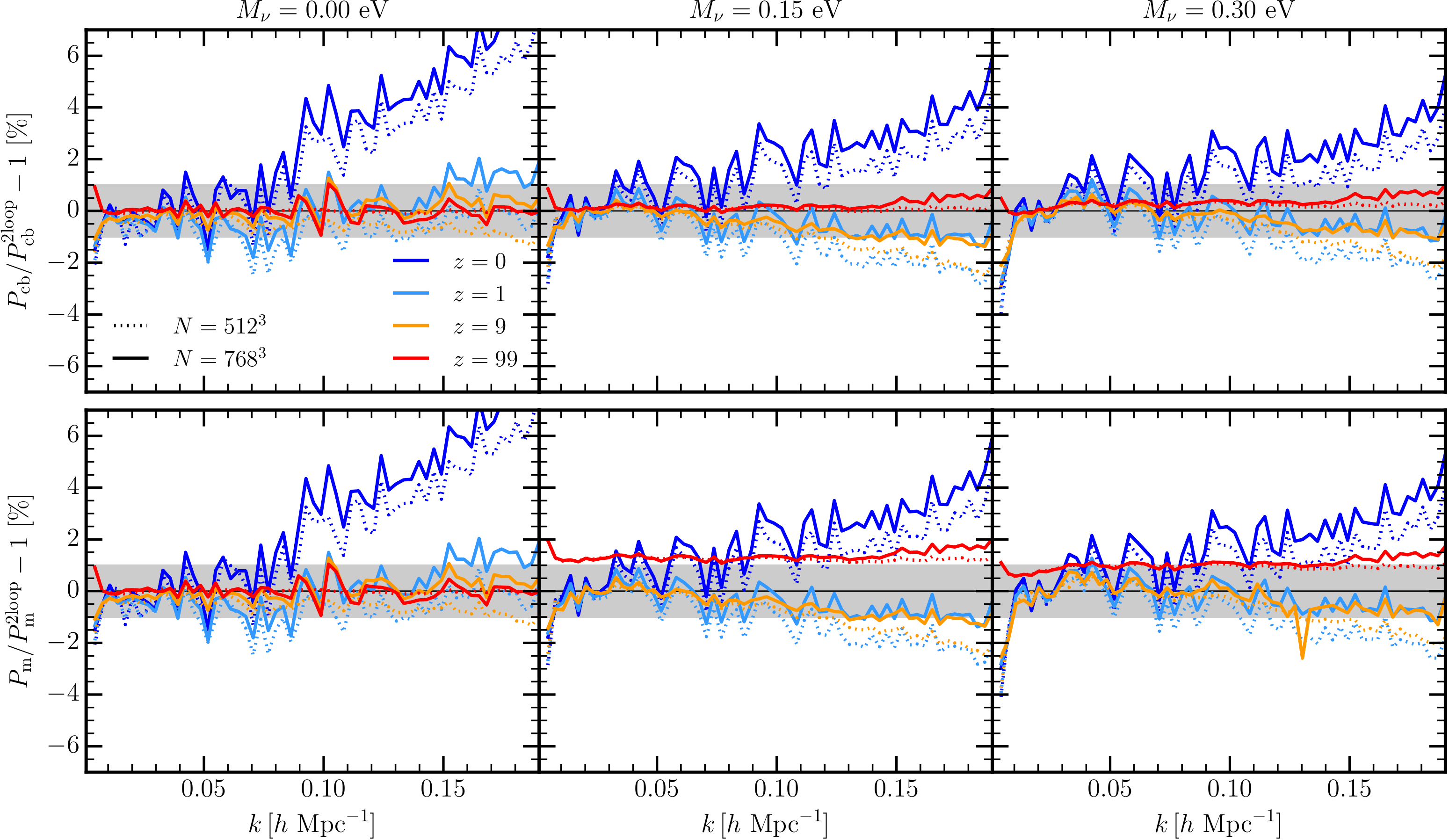}
\caption{Ratio of the power spectra measured in the simulations to nonlinear predictions for $\Lambda$CDM (left panels) and massive neutrinos, $M_\nu=0.15$ eV (middle panels) and $M_\nu=0.3$ eV (right panels). Nonlinear predictions are computed from the linear outputs of \camb{}, applying 2-loop corrections computed with the code \texttt{RegPT} for the CDM+baryon component only. The top panels show the power spectra of the cold dark matter component, while the bottom ones show the total matter power spectra. Colours from blue to red correspond to different redshifts, namely $z=0,1,9,99$. Solid lines refer to the simulations with $N=2 \times 768^3$ particles, while dotted lines to the simulation with $N= 512^3$ particles. }
\label{fig:sim_resolution}
\end{figure*}

Initial conditions are generated at $z=99$ by displacing and assigning peculiar velocities to particles, that initially are located in a regular grid, using the Zel'dovich approximation. For neutrino particles we also add a thermal velocity component. The amplitude of the thermal velocities is determined by randomly sampling the Fermi-Dirac distribution of the corresponding model while the direction is taken randomly within the sphere. Thermal velocities dominate neutrino dynamics during the first time-steps, having a dispersion roughly five orders of magnitude larger than the dispersion of peculiar neutrino velocities. Instead of sampling the modes amplitude in Fourier-space using a Rayleigh distribution (as in a Gaussian distribution), we collapse the distribution to its mean value. It can be shown that a simulation run with the initial conditions generated in that way will have the correct 2-point statistics with a lower variance \citep{AnguloPontzen2016}.

The displacements and peculiar velocities are computed taking into account the scale-dependent growth factor and growth rate using the procedure described in the previous section. We have modified the {\sc N-GenIC} code to achieve this. In the simulations with massive neutrinos we have generated the initial conditions for the scenarios S4 and S5 (see table \ref{tab:sims}). The simulations have been run using a tabulated Hubble function, that is different for each model and scenario, that controls the time evolution of the background in the simulations.

In Fig.~\ref{fig:sim_resolution} we show the ratio of the measured power spectra, $P_{m}(k)$ and $P_{cb}(k)$, to the corresponding nonlinear predictions for $M_\nu=0$ ($\Lambda$CDM), $0.15$ and $0.3$ eV. Nonlinear predictions are obtained with the code \texttt{RegPT} \citep{TaruyaEtal2012} implementing the multipoint propagator expansion of \citet{BernardeauCrocceScoccimarro2008}. We consider, in particular, the 2-loop approximation. In the case of massive neutrinos, we follow \citet{CastorinaEtal2015} and consider non linear predictions \textit{only} for the CDM+baryon component, since both the auto power spectrum of neutrinos and the cross power spectrum of cold matter and neutrinos can safely be described with linear theory at all redshifts. We find an agreement at the 1\% level between the output of the simulation and the predicted power spectra at all redshifts, except at $z=0$, for the simulations with $N=768^3$ particles (solid curves in the figure). At $z=0$, 2-loop regularised perturbation theory is known to underestimate the nonlinear power spectrum \citep{TaruyaEtal2012}. The agreement is worsen at small scales to a $\sim 2\%$ for the simulations with $N=512^3$ particles (dotted curves).

These results show that we are able to recover the same level of agreement in the massive neutrino case between simulations and predictions as in the massless neutrino scenario at low redshift. The residual differences between the two models can be ascribed to two effects. The first is the lower nonlinear evolution in the massive neutrino case (and therefore lower discrepancy with the nonlinear prediction). This is due to the lower amplitude of matter fluctuations of the neutrino model. The second is a larger resolution effect in the massive neutrino simulations: for higher resolution simulations as those of \citet{CastorinaEtal2015} this discrepancy disappears at mildly nonlinear scales.

The excess of power with respect to the predicted power spectrum that appears at the initial redshift in the first $k$-bin is expected from the Newtonian approximation assumed in our rescaling, since we are completely neglecting radiation (photon) perturbations. This approximation, being consistent with the simulation dynamics, allows us to recover the desired low-redshift power spectrum, that otherwise, according to Fig. \ref{fig:norm}, would be suppressed by $5-6\%$ on near-horizon scales. However, given that the size of the simulation box is close to the scale of the horizon at the initial redshift, we are not able to study this feature on a significant range of scales.

A peculiar feature of Fig.~\ref{fig:sim_resolution} is the behaviour of the total matter power spectrum at the initial redshift of the simulation. While the CDM power spectrum agrees with \camb{} at $z_i=99$, the total matter one shows an almost constant discrepancy of about 1\%. This is due to the fact that the simulation considers neutrinos as a non-relativistic species whose relative density with respect to CDM does not evolve, $\fnu = const$. On the contrary, in the Boltzmann code the evolution of neutrino density is \textit{properly} taken into account and the time-dependent $\fnu(z)$ is considered at each redshift. Moreover, the error in the case with $M_\nu=0.15$ eV is slightly larger than for $M_\nu=0.3$ eV, since neglecting the neutrino relativistic tail has a greater effect for lighter neutrinos. We checked that, by combining $P_{cb}$ and $P_\nu$ weighted with the same neutrino fraction $\fnu(z)$ used in \camb{} (or \class{}), we are able to recover the same level of agreement with the Boltzmann solution for the total matter power spectrum as for the CDM. Nonetheless, the total matter in the simulation is defined with a constant neutrino to CDM fraction and, therefore, intrinsically exhibits a discrepancy, at high redshift, with respect to the output of a Boltzmann code. This feature is avoided is grid-based and hybrid simulation codes, and could in principle be avoided also in particle-based codes by implementing a time-dependent neutrino particle mass. Still, even in our case, the agreement at low redshift is comparable to the $\Lambda$CDM case, being $\fnu = const$ a good approximation at late times. 

For the two massive neutrino simulations that we run, as mentioned in the above paragraphs, we chose to compute the rescaling in two different ways: in one case we implemented the features of scenario S4 (see section \ref{sec:ICs}), where neutrinos are treated as a non-relativistic species when computing the gravitational potential but the Hubble function is computed correctly. In the other case (S5), instead, neutrinos are treated as non-relativistic particles both in the dynamics and in the background. We therefore obtained two different initial power spectra, with case S5 exhibiting a slightly larger discrepancy with respect to \camb{} at the initial redshift, because of the approximation assumed in the Hubble function. Such difference at the initial redshifts amounts to $\sim 0.2\%$ at $k\sim0.1h~\mathrm{Mpc}^{-1}$, while for scenario S4 we have an agreement of $\sim0.05\%$ at the same scale. The two power spectra have then been evolved in the simulation, each with its own tabulated Hubble function. We found that these two methods give very similar results, both with $M_\nu=0.15$ and $M_\nu=0.3$ eV, recovering at low redshift the same agreement with the predicted power spectrum. As a matter of fact, the $z=0$ power spectra from the simulations run in scenario S4 and S5 differ by less that $0.05\%$ between each other for both neutrino masses.


\section{Conclusions}
\label{sec:conclusions}

Numerical N-body simulations are nowadays an indispensable tool in the study of the large-scale distribution of matter and galaxies in the Universe. While they capture the complex nonlinear evolution of structures that characterises gravitational instability, they nevertheless employ several approximations that might affect our ability to predict key quantities, like the matter power spectrum, at percent level accuracy. The necessity to ensure highly accurate numerical simulation for the next generation of galaxy and weak lensing surveys is indeed a quite relevant challenge for contemporary cosmology \citep[see \eg{}][]{SchneiderEtal2015,GarrisonEtal2016}.

In this work we quantified the errors resulting from some of the approximations that numerical simulations typically employ, in some cases necessarily so, with a specific attention to models including a massive neutrino component. Until recently, these scenarios have often been considered, rather oddly, as Beyond-the-Standard-Model, also because of the significant complications that their correct description would require. However, evidence for a neutrino mass, albeit small, is now beyond dispute and the effect of the latter on the matter power spectrum is quite significant. It is therefore crucial to look with renewed attention at these cosmologies for essentially two different reasons. First, an accurate modelling of neutrino effects is necessary to avoid potential systematic errors in the detection of possible, unexpected dark energy effects motivating a great part of current observational efforts in cosmology. Secondly, cosmological observations provide an upper limit to the neutrino mass now beyond reach for laboratory experiments and are potentially capable to provide us with a determination of the mass itself in the future.  

As numerical simulations work within the Newtonian approximation, completely neglecting radiation (in particular photon) perturbations, the set-up of the initial conditions (and of the subsequent expansion history) cannot be given by a direct match of the initial density and particle distribution to the linear prediction of a fully-relativistic Boltzmann code, without introducing systematic errors on the scale of the horizon. In fact, in simulations of $\Lambda$CDM cosmologies, the desired linear power spectrum at low redshift (where simulations are expected to provide the correct nonlinear evolution) is typically {\em rescaled} to the initial redshift within the same, Newtonian approximation assumed by the simulation dynamics, consistently neglecting photon perturbations in the rescaling procedure.

The main result of this work consists in showing how a similar procedure can be extended to simulations of massive neutrino cosmologies. To this end, we have adopted a two-fluid approximation to describe the linear evolution of the coupled cold matter and neutrino perturbations, consistently with the approximations adopted by the N-body simulations.

Before testing our method on particle-based N-body simulations, however, we have exploited the flexibility of the two-fluid differential equations to quantify the systematic errors on the linear evolution of perturbations that can result from possible choices in the definition of the initial particle positions and velocities, in addition to the expansion history provided to the N-body code. To this end we have assumed that the statistical properties of the initial density field are provided by a Boltzmann code at the reference initial redshift $z_i=99$. We have then evaluated the error on the $z=0$ linear power spectrum resulting from possible discontinuities introduced at $z_i$ or approximations adopted for the following evolution.

We have found, for instance, that the greatest effect is due to neglecting the radiation contribution in the computation of the expansion rate and, at the same time, using a scale-independent growth rate $f=\Omega_m^{0.55}$. This scenario alone accounts for a nearly $4 \%$ mismatch with respect to the reference $z=0$ power spectrum if the assumed initial redshift is $z_i=100$ (it decreases to a still significant $2\%$ for $z_i=50$). This mismatch amounts instead to a $1.5-2\%$ level when, still neglecting radiation, we use the correct, scale-dependent growth rate in the initial conditions. When radiation is included in the background, the largest error comes from the approximated value $f=\Omega_m^{0.55}$, which results in a $0.3\%$ mismatch in a cosmology without massive neutrinos and can reach $\sim 1\%$ level when massive neutrinos are considered. On the other hand, including radiation and using the asymptotic value of the growth rate (see case S3b of Tab \ref{tab:sims}) significantly alleviates the systematic error, which becomes negligible on small scales. Nonetheless, on scales approaching the horizon, the mismatch on the evolved power spectra still remains critical.

Other approximations are inevitably related to the way massive neutrinos perturbations are described in the simulation. For instance, particle-based simulations consider neutrinos as a fully non-relativistic species (described by particles with fixed mass) at {\em all} redshifts. If we do not correct for this in the initial conditions, the resulting error is of order $0.1\%$ on the CDM low-redshift power spectrum, but can approach $0.8-0.9\%$ on the total matter power spectrum. Clearly, a significant systematic error can be the consequence of the combination of several of such approximations, even when their individual effect is in principle negligible.

This exercise allowed us to identify the optimal choices for setting up the initial conditions, in order to keep such systematic errors under control. In addition, we have used the Newtonian, two-fluid approximation for matter and neutrino perturbations  to perform the proper, {\em scale-dependent} rescaling of the input $z=0$ linear power spectra to $z_i$, in order to avoid all errors related to approximations to the expansion history (\eg{} no radiation).

We have tested this method to compute the displacements and velocities of particles in the initial conditions of cosmological simulations, both with and without massive neutrinos. We have found that, regardless of the choice of neutrino total mass, we are able to recover subpercent agreement between the total matter and cold dark matter power spectra measured in the simulation at low redshift ($z\lesssim 10$, and in the linear regime) and the Boltzmann, linear prediction. A larger discrepancy at very high redshift is expected from the approximations assumed, unavoidable for standard N-body simulations.

We have implemented the methods presented in this work in the public code \texttt{REPS} --  rescaled power spectra for initial conditions with massive neutrinos. We stress once more that massive neutrino models do represent the current ``standard'' cosmological models. Properly quantifying and correcting systematic errors as we have done in this work is crucial to ensure that N-body simulations will reach the accuracy required by dark energy studies.

\subsection*{Acknowledgements} 
LG, CC and JB acknowledge financial support from the European Research Council through the Darklight Advanced Research Grant (n. 291521). FVN is supported by the ERC Starting Grant ``cosmoIGM'' and partially supported by INFN IS PD51 ``INDARK''. We thank Simeon Bird, Emanuele Castorina, Enea Di Dio and Matteo Viel for useful conversations. N-body simulations have been run in the Zefiro cluster (Pisa, Italy).
\bibliography{Bibliography}

\end{document}